\documentstyle[epsf]{article}
\begin{document}
\begin{center}
{\large\bf 0n the Fluctuation Law(s)\\ for Hamiltonian systems (with equilibrium steady state):\\
           A Comment on cond--mat/0008421}\\[2mm]

                   Boris Chirikov and Oleg Zhirov\\
{\it Budker Institute of Nuclear Physics \\
        630090 Novosibirsk, Russia}\\[1mm]
        chirikov @ inp.nsk.su\\
        zhirov @ inp.nsk.su\\[5mm]
\end{center}

\baselineskip=15pt

\hspace*{1cm}

A generalization of the fluctuation law (FL) ("theorem"), formulated in 1993
by Evans, Cohen and Morriss for a nonequilibrium steady state,
on the chaotic Hamiltonian systems with equilibrium steady state in recent
publication by Evans, Searles and Mittag (cond--mat/0008421) is briefly
discussed. We argue that the physical meaning of this law, as presented
in the latter publication, is qualitatively different from the original one.
Namely, the original FL concerns the {\it local} (in time) fluctuations
with an intriguing result: a high probability for the "violation"
of the Second Law. Instead, the new law describes the {\it global}
fluctuations for which this remarkable
unexpected phenomenon is absent or hidden.
We compare both types of fluctuations in both classes of Hamiltonian systems,
and discuss remarkable similarities
as well as the interesting distinctions.

\hspace*{1cm}

The "Fluctuation
Theorem" has been first obtained by Evans, Cohen and Morriss
\cite{1} for a particular example of the nonequilibrium steady
state, using both the theory as well as numerics. For our
purposes it can be represented in the form:
$$
   \ln{\left({p(\Delta S)\over p(-\Delta S)}\right)}=
   F\cdot\Delta S\,, \qquad
   F = {2\langle\Delta S\rangle\over \sigma^2}
   \eqno(1)
$$
Here $p(\Delta S)$ is the probability of entropy
(or entropy--like quantity as in \cite{1,2}) change $\Delta S$ in the ensemble
of trajectory segments of a fixed (appropriately scaled) duration $\tau$ for the
mean change $\langle\Delta S\rangle\ >\ 0$ and variance
$\sigma^2$, and $F$ the fluctuation parameter usually
taken to be unity ($F=1$).
We call this type of fluctuations the {\it local}
fluctuations.

By itself, the relation (1) is but a specific reduced
representation of
the normal probabilistic law, the Gaussian distribution,
in a suitable random variable ($\Delta S$):
$$
   p(\Delta S) = {1\over \sqrt{2\pi\sigma^2}}\cdot
   \exp{\left(-{(\Delta S - \langle\Delta S\rangle )^2\over
   2\sigma^2}\right)} \eqno (2)
$$
shifted with respect to $\Delta S = 0$ due to the permanent entropy production at a constant rate in the
nonequilibrium steady state. The FL (1) immediately follows
from the normal law (2) but not vice versa.
Notice also that this distribution
is not universal, yet it is rather typical indeed.
However, the surprise (to many) was in that the probability
of {\it negative} ("abnormal") entropy change $\Delta S < 0$
(without time reversal!)
is generally not small at all reaching 50\% for sufficiently short $\tau$.
That is every second change may be abnormal !?

Implicitly, all that is contained in the well developed
statistical theory (see, e.g., \cite{3}, Section 20,
and \cite{4}).
Nevertheless, the first
direct observation of this phenomenon in a nonequilibrium
steady state \cite{1} has so much impressed
the authors that they even entitled the paper "Probability of
Second Law violations in shearing steady state". In fact, this is simply
a sort of peculiar fluctuations discussed in
\cite{5}.

In our opinion, the main
lesson one should learn from the FL is that the entropy
evolution is generally {\it nonmonotonic} contrary to a common belief,
still now.
The origin of this confusion is, perhaps,
in traditional conception of the fluctuations as a
charcteristic on the microscopic scale well separated from
a much larger macroscopic scale with its averaged quantities
like the entropy production rate, for example.

In equilibrium steady state the macroscopic scale with the mean rate
$\langle\Delta S\rangle =0$ traditionally seems to be irrelevant
with the entropy trivially conserved. However,
in nonequilibrium
steady state the macroscopic scale is represented by a finite
rate $\langle\Delta S\rangle\ >\ 0$, yet the "microscopic"
scale of the fluctuations may be well comparable with, and even
exceed, the former.

The border of nonmonotonic behavior is at $\Delta S = 0$
(no entropy rise at all) which corresponds to the probability
of "abnormal" entropy changes $\Delta S < 0$
$$
   P_{ab} = \int_{-\infty}^0\,p(s)\,ds\,, \qquad
   C={\langle\Delta S\rangle\over \sigma} \eqno(3)
$$
Here $C$ is a new parameter for the normal/abnormal crossover
in the entropy variation sign at $|C|=C_{cro}\sim 1$ when
the probability $P_{ab}$ is large.

If a finite--dimensional Hamiltonian system admits the (stable)
statistical equilibrium (as is the case in \cite{2}) the
overall ($t\to\infty$) average entropy rate
$\langle\Delta S\rangle\ = 0$ for any $\tau$. However, on a
finite time scale $t_R$ of a nonequilibrium relaxation to the
equilibrium the local $\langle\Delta S\rangle\ >\ 0$ as in the
nonequilibrium steady state, but temporally. On this time scale
the local fluctuations are expected \cite{5} to obey the law
similar to the nonequilibrium FL provided $\tau\ll \tau_R$.
However, the properties of fluctuations in such a system,
studied in \cite{2}, turned out to be qualitatively different.
Particularly, the celebrated phenomenon of the abnormal
fluctuations completely disappeared (?).
In our understanding, this crucial change is caused by a
different type of the fluctuations studied in \cite{2}:
instead to fix $\tau\ll \tau_R$ the fluctuations as a function
of time $S(t)=A(t)$ (in \cite{2}) were considered.
Such global fluctuations were also studied in a nonequilibrium
steady state \cite{5}, and found to be rather different, indeed.
Particularly, the crossover parameter
$$
   C(t) = {\langle S(t)\rangle\over \sigma} \eqno(4)
$$
depends now on the motion time, which is a dynamical variable,
rather then on the trajectory segment length $\tau$, which is
a parameter of the problem.

It is useful to compare both types of fluctuations for both
classes of dynamical systems in nonequilibrium
(far--from--equilibrium) states.
For the sake of brevity, let us
call the new class \cite{2} the stable equilibrium class (SEQ)
while the original class (like in \cite{1}) the equilibriumfree,
or no--equilibrium one (NEQ). Indeed, the systems with and
without the equilibrium steady state have rather different
statistical properties of the motion \cite{5}. Notice that,
from the physical point of view, both classes are Hamiltonian.
In our opinion, the simple models of NEQ with a reversible
"friction", being formally dissipative, are physically interesting as far as they simulate
the real Hamiltonian infinite--dimensional thermostat.

The local fluctuations in the nonequilibrium {\it steady state}
of NEQ can be determined for any segment length $\tau$, and
do not depend on the motion time $t$. Same fluctuations in
SEQ for a {\it transient, finite--time} nonequilibrium relaxation require the
additional condition $\tau\ll \tau_R$ to be measured, and they
do depend on the motion time. For $t\gg t_R$ the fluctuations
as well as other statistical properties become equilibrium ones.
In this case the crossover parameter $C(t)\to 0$ together with
the average $\langle\Delta S\rangle\to 0$, and hence half of
changes $\Delta S$ become "abnormal" as it should be in the statistical
equilibrium.

Unlike this, the global fluctuations in NEQ do always depend
on time $t$, and parameter\\ $C(t)\to\infty$ due to
$\langle S(t)\rangle\to\infty$. In the limit $t\to\infty$
this implies {\it zero} probability of abnormal $S(t)<S(0)$
(see \cite{5}). Same fluctuations in SEQ are not exactly the same,
and have an interesting peculiarity found in \cite{2}:
the {\it finite} asymptotic probability of abnormal fluctuations.
Moreover, this asymptotic value is of the order
of the probability for a spontaneous
fluctuation of 
the size comparable with the initial
nonequilibrium state of the system which would subsequently relax
back to the equilibrium, and so forth ad infinitum. This is not a completely new
phenomenon but, apparently, less known one (for discussion
see \cite{5} and references therein).

In conclusion, let us mention that we would expect the
{\it rise} of big spontaneous fluctuations, implicitly observed
in \cite{2}, to obey the law called in \cite{2}
"Anti--Fluctuation Theorem" but {\it without} time reversal.

\end{document}